\documentclass[aps,pr,superscriptaddress,preprintnumbers,nofootinbib,10pt]{revtex4-2}
\usepackage{multirow}
\usepackage{amsmath}
\usepackage{amssymb}
\usepackage[dvipdf,dvips]{graphicx}
\usepackage{color}
\usepackage{hyperref}
\usepackage{url}
\usepackage{slashed}
\usepackage{subfigure}
\usepackage[usenames,dvipsnames]{xcolor}
\usepackage{amsmath}
\usepackage{amsfonts}
\usepackage{float} 
\usepackage{amssymb}
\usepackage{epsfig}
\usepackage{graphics}
\usepackage{euscript}
\usepackage{slashed}
\usepackage{epstopdf}
\usepackage[utf8]{inputenc}
\allowdisplaybreaks
\usepackage[normalem]{ulem}
\usepackage{pifont}
\usepackage{dsfont}
\usepackage{MnSymbol}
\usepackage{verbatim}
\usepackage{graphicx}
\usepackage{latexsym}
\usepackage{tikz-feynman}
\usepackage{bbold}
\usepackage{lineno}
\usepackage{enumitem}

\begin{document}

	\title{{\bf   A study of spin $1$  Unruh-De Witt  detectors}}

	\author{F. M. Guedes} \email{fmqguedes@gmail.com} \affiliation{UERJ $–$ Universidade do Estado do Rio de Janeiro,	Instituto de Física $–$ Departamento de Física Teórica $–$ Rua São Francisco Xavier 524, 20550-013, Maracanã, Rio de Janeiro, Brazil}
	
	\author{M. S.  Guimaraes}\email{msguimaraes@uerj.br} \affiliation{UERJ $–$ Universidade do Estado do Rio de Janeiro,	Instituto de Física $–$ Departamento de Física Teórica $–$ Rua São Francisco Xavier 524, 20550-013, Maracanã, Rio de Janeiro, Brazil}
	
	\author{I. Roditi} \email{roditi@cbpf.br} \affiliation{CBPF $-$ Centro Brasileiro de Pesquisas Físicas, Rua Dr. Xavier Sigaud 150, 22290-180, Rio de Janeiro, Brazil}
	
	\author{S. P. Sorella} \email{silvio.sorella@gmail.com} \affiliation{UERJ $–$ Universidade do Estado do Rio de Janeiro,	Instituto de Física $–$ Departamento de Física Teórica $–$ Rua São Francisco Xavier 524, 20550-013, Maracanã, Rio de Janeiro, Brazil}

	\begin{abstract}

	A study of the spin 1 Unruh-De Witt detectors interacting with a relativistic scalar quantum field is presented. After tracing out the field modes, the resulting density matrix for a bipartite qutrit system is employed to investigate the violation of the Bell-CHSH inequality. Unlike the case of spin $1/2$, for which the effects of the quantum field result in a decreasing of the size of violation, in the case of spin $1$ both decreasing and increasing of the violation may occur. This effect is ascribed to the fact that Tsirelson's bound is not saturated in the case of qutrits. 
	
	\end{abstract}
	
	\maketitle

\section{Introduction}	
\label{intro}

\noindent The Unruh-De Witt (UDW) detectors are useful models,  broadly employed in the study of relativistic quantum information, see  \cite{Reznik:2002fz,Tjoa:2022vnq,Tjoa:2022lel,Lima:2023pyt,Barbado:2020snx,Foo:2020xqn,Verdon-Akzam:2015tma}. \\\\In this work, we shall employ  spin $1$ Unruh-De Witt detectors to investigate the  effects of a quantum relativistic scalar field on the Bell-CHSH inequality \cite{Bell64, CHSH69}, following the setup already outlined for spin $1/2$ detectors  \cite{spinhalf24}. More precisely, we consider the interaction of a pair of  qutrits with a real Klein-Gordon field in Minkowski spacetime, by taking as initial field configuration the vacuum state  $| 0 \rangle$. \\\\{ One starts with the density matrix corresponding to the state  
\begin{equation} 
| \psi \rangle_{AB} \otimes |0\rangle \;, \label{sti}
\end{equation}
where 
\begin{align}
   \vert \psi \rangle_{AB} = \frac{1}{\sqrt{3}} \left ( \vert 1 \rangle_{A} \otimes\vert -1 \rangle_{B} - \vert 0 \rangle_{A} \otimes\vert 0 \rangle_{B} + \vert -1 \rangle_{A} \otimes\vert 1 \rangle_{B} \right ),
    \label{stateqtrit}
\end{align}
is the maximally entangled singlet state of two qutrits. As it is customary, $(A,B)$ refer to Alice and Bob, respectively. \\\\One let evolve the density matrix by means of the unitary operator corresponding to the dephasing channel. The resulting asymptotic density matrix is employed to study the effects arising from the presence of the scalar field on the violations of the Bell-CHSH inequality. Though, one has to remember that the operators $(A,A')$ and $(B,B')$ entering the Bell-CHSH correlator ${\cal C}$, Eq.\eqref{Cc}, are required to fulfill the conditions \eqref{Bop}, implying that $(A,A')$ and $(B,B')$ have to be space-like, in agreement with relativistic causality. This feature is taken into account by employing the right and left wedges, ${\cal W}^R$, ${\cal W}^L$: 
\begin{equation} 
{\cal W}^R = \{ x; x > |t| \} \;, \qquad {\cal W}^L = \{ x;, -x> |t| \} \;.\label{wedges}
\end{equation}
Regions contained in ${\cal W}^R$ are space-like with respect to regions of ${\cal W}^L$.  \\\\Moreover, as one learns from  \cite{SW1,Summers87b}, the use of the wedges regions $({\cal W}^R$, ${\cal W}^L)$ enables us to employ the results about the nature of the vacuum state $|0\rangle$. It has been established \cite{SW1,Summers87b} that the vacuum $|0\rangle$ is highly entangled, exhibiting maximal violation of the Quantum Field Theory formulation of the Bell-CHSH inequality for regions belonging to  the wedges. As such, the state \eqref{sti} looks ideal for a study of the effects of the quantum field on the violation of the Bell-CHSH inequality for the qutrits system.  \\\\Nevertheless, as we shall see, there are remarkable differences between the spin $1/2$ and the spin $1$ cases. As far as the Bell-CHSH inequality is concerned, for spin $1/2$, the effects induced by the quantum field result in a decreasing of the size of the violation, due to the fact that the Tsirelson bound \cite{TSI}, {\it i.e.} $2 \sqrt{2}$, is already attained in the absence of the field. As the Tsirelson bound is the maximum allowed value for the violation, one can easily figure out that the presence of a quantum field can only induce a decreasing of the size of the violation, see \cite{spinhalf24} for more details. Instead, in the case of spin $1$, the situation looks rather different. Here, it is known that Tsirelson's bound is never attained \cite{Gisin,Peruzzo:2023nrr}. The maximum value for the Bell-CHSH inequality is approximately $2.55$.  As such, depending on the choice of the parameters, the effects of the quantum field may give rise either to a decreasing or to an increasing of the violation, while remaining compatible with Tsirelson's bound $2\sqrt{2}$.  In the case of a decreasing one has a degradation  \cite{Lin:2008jj} of the entanglement properties of the initial state, while in the case of an increasing of the violation one might speak of extraction of entanglement \cite{Reznik:2002fz,Tjoa:2022vnq,Tjoa:2022lel,Lima:2023pyt,Barbado:2020snx,Foo:2020xqn,Verdon-Akzam:2015tma}. }\\\\This work is organized as follows. In Sec.\eqref{evaluation rho AB}, we evaluate the  qutrit density matrix by considering the dephasing coupling regime. In Sec.\eqref{TTW} we provide an overview of the fundamental characteristics of the Weyl operators $W_{f_j}$ 
\begin{equation} 
W_{f_j} = e^{i {\varphi}(f_j) } \;, \qquad j=A,B \;, 
\label{Weyli}
\end{equation}
and their von Neumann algebra, introducing key concepts that will be employed throughout this study. In Sec.\eqref{BCHSH}, we discuss how the effects  of the quantum field $\varphi $ on the violation of the Bell-CHSH inequality, which can be obtained in closed form by using the powerful modular theory of Tomita-Takesaki \cite{Witten18,Bratteli97,SW1,Summers87b,Weyl23}. Notably, it turns out that the violation of the Bell-CHSH inequality exhibits both an increasing and a decreasing behavior as compared to the case in which the field $\varphi$ is absent, Sec.\eqref{conclusion} collects our conclusion. 

\subsection{Preliminaries} 
\noindent For the initial density matrix we have
\begin{align}
    \rho_{AB\varphi}(0) = \rho_{AB}(0)  \otimes  |0 \rangle    \langle 0| \;, \label{ddo}
\end{align}
where 
\begin{equation} 
\rho_{AB}(0)= \vert \psi  \rangle_{AB}\; _{AB} \langle \psi | \;. \label{qtrho}
\end{equation}
The time evolution of   $\rho_{AB\varphi}(0)$  is governed by the unitary operator 
\begin{align}
    \mathcal{U} = e^{-i [J_{A}^{z} \otimes \varphi(f_{A}) + J_{B}^{z} \otimes \varphi(f_{B})]} \;,
    \label{Unit operator}
\end{align}
where the operator $J^{z}$  corresponds to the component of spin along the $z$-axis, and  $\varphi(f_j)$, $j= A,B$, is the smeared field \cite{Haag92}:
\begin{equation} 
\varphi(f_j) = \int d^4x \; \varphi(x) f_j(x) \;, \qquad j=A, B \;, \label{smvphi}
\end{equation}
{ where $f_j(x)$ are smooth test functions with compact support\footnote{A smooth function f with compact support ${\cal M} \in {\mathbb{R}^4 }$ is a continuous infinitely differentiable function vanishing outside the region ${\cal M}$.},  $f_j(x) \in {\cal C}_{0}^{\infty}({\mathbb{R}}^4)$. As mentioned before, the support of Alice's test function $f_A(x)$ is an open region ${\cal O} \in {\cal W}^R$. Relying thus on the powerful Tomita-Takesaki modular theory for von Neumann algebras \cite{Witten18,Bratteli97,SW1,Summers87b,Weyl23},  Bob's test function $f_B(x)$  will be supported in the causal complement ${\cal O}'$ of ${\cal O}$, located in ${\cal W}^L$. The norms and the Lorentz invariant inner products of $(f_A,f_B)$ are also determined by the properties of the modular theory, as given in Eqs.\eqref{sfl}, see the review \cite{Guido:2008jk} for a detailed account. The role of the test functions   $f_j$ is that of localizing the quantum field in the regions mentioned above.}\\\\For the quantum field $\varphi$, one writes 
\begin{eqnarray}
\varphi(x) & = & \int \frac{d^3p}{(2\pi)^3}\frac{1}{2\omega_p} \left( e^{-ipx} a_p + e^{ipx} a^{\dagger}_p \right) \;, \qquad \omega_p=\sqrt{{\vec p}^2+m^2} \;, \qquad px=p_0 x_0 - \vec{p} \cdot \vec{x} \nonumber \\
&\;& [ a_p , a^{\dagger}_q]  =  (2\pi)^3 \; (2\omega_p) \;\delta^{3}(p-q) \;, \qquad [a_p,a_q]= 0\;. \label{qvft}
\end{eqnarray}
{ Let us proceed by providing the derivation of the unitary evolution operator of Eq.\eqref{Unit operator}. One starts with the Hamiltonian 
\begin{equation}
H = H_0 + H_I(t) \;, \label{hm}
\end{equation}
where $H_0$ stands for the free Hamiltonian 
\begin{equation} 
H_0 = \sigma (J^A_z \otimes \mathbb{1}^B + J^B_z \otimes \mathbb{1}^A ) + \int d\mu(p)\; \omega_p a^{\dagger}_p a_p \;,\qquad d\mu(p)=  \frac{d^3p}{(2\pi)^3}\frac{1}{2\omega_p} \;, \label{ho} 
\end{equation}
and $H_I(t)$ is the interaction term:
\begin{equation} 
H_I(t) = J^A_z \int d^3x \varphi(x,0) f_A(x,t) + J^B_z  \int d^3x \varphi(x,0) f_B(x,t) \;. \label{hi} 
\end{equation}
Notice that 
\begin{equation} 
H_0 | \psi \rangle_{AB} \otimes |0\rangle =0 \;. \label{zz}
\end{equation} 
For the evolution operator in the interaction representation, we have 
\begin{equation} 
{\cal U}(t) = T_t e^{-i \int^t d\tau H_I(\tau) } \;, \label{evu}
\end{equation}
where $T_t$ is the time ordering. In order to work out expression \eqref{evu}, one makes use of the Magnus formula \cite{Tjoa:2022vnq,Tjoa:2022lel}, summarized as 
\begin{equation} 
T_t \; e^{\int^{t} d\tau A(\tau) }= e^{\Omega(t)} \;, \qquad \Omega(t) = \sum_{n=1}^{\infty} \Omega_n(t)   \label{mag}
\end{equation} 
with
\begin{eqnarray} 
\Omega_1(t) & = & \int^t d\tau A(\tau) \nonumber \\
\Omega_2(t) &= &\frac{1}{2} \int^t d\tau_1 \int^{\tau_1} d\tau_2\; [A(\tau_1), A(\tau_2)] \nonumber \\
\Omega_j & = & {\rm higher \; order \; commutators } \;\;\;\; j\ge 3 \;. \label{expm}
\end{eqnarray} 
We remind now that the field commutator $[\varphi(x), \varphi(y)]$ is a $c$-number. As a consequence, $\Omega_j=0, i\ge3$, while $\Omega_2$ yields an irrelevant phase. Therefore, up to an irrelevant phase, for the evolution operator ${\cal U}(t)$, one gets 
\begin{equation} 
{\cal U}(t) = e^{-i \int^t d\tau H_I(\tau) } \;. \label{uu} 
\end{equation} 
Therefore, in the large asymptotic time, $t \rightarrow \infty$, equation  \eqref{Unit operator} follows, namely, 
at  large time, the density matrix is written as
\begin{align}
    {\rho}_{AB\varphi} = {\rho}_{AB\varphi}(t \rightarrow \infty) = \mathcal{U} \; \rho_{AB \varphi}(0)  \; \mathcal{U}^{\dagger}\;. \label{ddt}
\end{align}
The subsequent stage involves deriving the density matrix ${\hat \rho}_{AB}$ for the  qutrit system through the process of tracing out the field modes:
\begin{equation} 
{\hat \rho}_{AB} = {\rm Tr}_{\varphi} (\rho_{AB\varphi}) \;. \label{trace}
\end{equation} 
Finally, once the density matrix ${\hat \rho}_{AB}$ is known, one is capable of evaluating the Bell-CHSH correlator 
\begin{equation}  
\langle \mathcal{C} \rangle  =  {\rm Tr}({\hat \rho}_{AB} \mathcal{C}) \;, \label{tracerhoC}
\end{equation}
where
\begin{equation}
\mathcal{C}  =  (A+A')\otimes B + (A-A')\otimes B' \;, \label{Cc}
\end{equation}
with $(A,A')$, $(B,B')$ being the Bell operators,  namely 
\begin{eqnarray} 
A & = & A^{\dagger} \;, \quad A'= A'^{\dagger}\;, \quad B= B^{\dagger} \;, \quad B'= B'^{\dagger} 
\nonumber \\
A^2 & = & A'^2 = B^2=B'^2=1  \nonumber \\
\left[A,B \right] & = & [A,B'] = [A',B] = [A', B'] =0  \;. \label{Bop}
\end{eqnarray}  
Concerning the commutators $[A,A']$ and $ [B,B']$, they can  be expressed in terms of the four Bell's parameters $(\alpha, \alpha',\beta, \beta')$, Eq.\eqref{bopp}, {\it i.e.}
\begin{equation} 
[A,A'] = 
\begin{pmatrix} 
2 \sin(\alpha-\alpha') & 0 & 0 \\
0 & 0 & 0 \\
0 & 0 & 2 \sin(\alpha'-\alpha) 
\end{pmatrix}
\label{amm}
\end{equation} 
and 
\begin{equation} 
[B,B'] = 
\begin{pmatrix} 
2 \sin(\beta'-\beta) & 0 & 0 \\
0 & 0 & 0 \\
0 & 0 & 2 \sin(\beta-\beta') 
\end{pmatrix}
\label{bmm}
\end{equation} 
The Bell-CHSH inequality is said to be violated whenever 
\begin{equation} 
2 < | \langle {\cal C} \rangle | \le 2 \sqrt{2} \;. \label{tsi}
\end{equation}
 
}

\section{Evaluation of the qutrit density matrix in the case of the dephasing coupling detectors} \label{evaluation rho AB}

\noindent We shall consider the density matrix ${\hat \rho}_{AB}$ in he so-called dephasing coupling  regime \cite{Tjoa:2022vnq,Tjoa:2022lel},  for which the evolution  operator is given by $\mathcal{U} = \mathcal{U}_{A} \otimes \mathcal{U}_{B}$, where the unitary operator for the detector $j=A,B$ is
\begin{align}
    \mathcal{U}_j = e^{-i J_{j}^{z} \otimes \varphi(f_{j})},
    \label{Unitary op j}
\end{align}
with the commutation relation
\begin{align}
\left [ \mathcal{U}_{A}, \mathcal{U}_B \right ] = 0. \label{uaub}
\end{align}
{ The above commutation relation follows from the fact that Alice's and Bob's test functions $(f_A,f_B)$ are space-like. This feature enables for several practical simplifications in the evaluation of the resulting density matrix for the qutrits system. }\\\\Using the algebra of the spin $1$  matrices, the expression \eqref{Unitary op j} can be written as 
\begin{align}
    \mathcal{U}_j ={ \mathbb{1}}_j - i  J_{j}^{z} s_{j} + (J_{j}^{z})^2 (c_{j} - 1),
\end{align}
where $c_{j} \equiv \cos \varphi (f_{j})$ and $s_{j} \equiv \sin \varphi (f_{j})$. With the initial density matrix $\rho_{AB\varphi}(0)$ given in Eq.\eqref{ddo}, its evolution is described as follows:
\begin{align}
    \rho_{AB \varphi} =&\left ( \mathcal{U}_{A} \otimes \mathcal{U}_{B} \right ) \rho_{AB\varphi}(0) \;(\mathcal{U}_{A}^{\dagger} \otimes \mathcal{U}_{B}^{\dagger}) \nonumber \\
    =& \left [ {\mathbb{1}}_A -i J_{A}^{z} s_A (J_{A}^{z})^2 (c_A -1 ) \right ] \otimes \left [{ \mathbb{1}}_B -i J_{B}^{z} s_B (J_{B}^{z})^2 (c_B -1 ) \right ] \rho_{AB}(0) \vert 0 \rangle \langle 0 \vert \nonumber \\
    & \times \left[ {\mathbb{1}}_A +i J_{A}^{z} s_A (J_{A}^{z})^2 (c_A -1 ) \right ] \otimes \left [ {\mathbb{1}}_B +i J_{B}^{z} s_B (J_{B}^{z})^2 (c_B -1 ) \right ]  \;.
    \label{rhoABphi dephasing}
\end{align}
Tracing over $\varphi$, we get a rather lengthy expression for $\hat{\rho}_{AB}$, namely 
\begin{align}
\hat{\rho}_{AB} = &  \rho_{AB}(0) + \rho_{AB}(0)(J_{B}^{z})^2 \langle c_B - 1 \rangle - \rho_{AB}(0) J_{A}^{z} \otimes J_{B}^{z} \langle s_A s_B \rangle + \rho_{AB}(0) (J_{A}^{z})^2 \langle c_A - 1 \rangle \nonumber \\
& + \rho_{AB}(0)(J_{A}^{z})^2 \otimes (J_{B}^{z})^2 \langle (c_A - 1)(c_B - 1) \rangle + (J_{B}^{z})^2 \rho_{AB}(0) J_{B}^{z} \langle s_B ^2 \rangle + (J_{B}^{z})^2  \rho_{AB}(0) J_{A}^{z} \langle s_B s_A \rangle \nonumber \\ & + J_{B}^{z} \rho_{AB}(0) J_{A}^{z} \otimes (J_{B}^{z})^2 \langle s_B s_A (c_B - 1 ) \rangle + J_{B}^{z} \rho_{AB}(0) (J_{A}^{z})^2 \otimes J_{B}^{z}  \langle s_B ^2 (c_A - 1 ) \rangle \nonumber \\
& + (J_{B}^{z})^2 \rho_{AB}(0) \langle c_B - 1 \rangle + (J_{B}^{z})^2 \rho_{AB}(0) (J_{B}^{z})^2 \langle (c_B - 1)^2 \rangle - (J_{B}^{z})^2 \rho_{AB}(0) J_z ^B \otimes J_z ^A \langle (c_B - 1) s_A s_B \rangle \nonumber \\
& + (J_{B}^{z})^2 \rho_{AB}(0) J_z ^A \langle (c_B - 1)(c_A - 1) \rangle + (J_{B}^{z})^2 \rho_{AB}(0) (J_{B}^{z})^2 \otimes (J_{A}^{z})^2 \langle (c_B - 1)^2 (c_A - 1) \rangle \nonumber \\
& + J_A ^z \rho_{AB}(0) J_B ^z \langle s_A s_B \rangle + J_A ^z \rho_{AB}(0) J_A ^z \langle s_A ^2 \rangle + J_A ^z \rho_{AB}(0) J_A ^z \otimes (J_B ^z)^2 \langle s_A ^2 (c_B -1) \rangle \nonumber \\
& + J_A ^z \rho_{AB}(0) (J_A ^z)^2 \otimes J_B ^z \langle s_A s_B (c_A -1) \rangle - J_A ^z \otimes J_B ^z \rho_{AB}(0) \langle s_A s_B \rangle - J_A ^z \otimes J_B ^z \rho_{AB}(0) (J_B ^z)^2 \langle s_A s_B (c_B -1 ) \rangle \nonumber \\
& + J_A ^z \otimes J_B ^z \rho_{AB}(0) J_A ^z \otimes J_B ^z \langle s_A ^2  s_B ^2 \rangle - J_A ^z \otimes J_B ^z \rho_{AB}(0) (J_A ^z)^2 \langle s_A   s_B (c_A -1) \rangle \nonumber \\
& - J_A ^z \otimes J_B ^z \rho_{AB}(0) (J_A ^z)^2 \otimes (J_B ^z)^2 \langle s_A   s_B (c_A -1)(c_B -1) \rangle + J_A ^z \otimes (J_B ^z)^2 \rho_{AB}(0) J_B ^z \langle s_A   s_B (c_B -1) \rangle \nonumber \\
& + J_A ^z \otimes (J_B ^z)^2 \rho_{AB}(0) J_A ^z  \langle s_A ^2 (c_B -1) \rangle + J_A ^z \otimes (J_B ^z)^2 \rho_{AB}(0) J_A ^z \otimes (J_B ^z)^2 \langle s_A ^2 (c_B -1)^2 \rangle \nonumber \\
& + J_A ^z \otimes (J_B ^z)^2 \rho_{AB}(0) (J_A ^z)^2 \otimes J_B ^z \langle s_A   s_B (c_B -1)(c_A -1)\rangle + (J_A ^z)^2 \rho_{AB}(0) \langle (c_A -1) \rangle \nonumber \\
& +  (J_A ^z)^2 \rho_{AB}(0) (J_B ^z)^2 \langle (c_A -1)(c_B -1) \rangle - (J_A ^z)^2 \rho_{AB}(0) J_A ^z \otimes J_B ^z \langle (c_A -1)s_A s_B \rangle \nonumber \\
& + (J_A ^z)^2 \rho_{AB}(0) (J_A ^z)^2 \langle (c_A -1)^2 \rangle + (J_A ^z)^2 \rho_{AB}(0) (J_A ^z)^2 \otimes (J_B ^z)^2 \langle (c_A -1)^2 (c_B -1) \rangle \nonumber \\
& + (J_A ^z)^2 \otimes J_B ^z \rho_{AB}(0) J_B ^z \langle (c_A -1) s_B ^2 \rangle + (J_A ^z)^2 \otimes J_B ^z \rho_{AB}(0) J_A ^z \langle (c_A -1) s_B s_A \rangle \nonumber \\
& + (J_A ^z)^2 \otimes  J_B ^z \rho_{AB}(0) J_A ^z \otimes (J_B ^z)^2 \langle (c_A -1)(c_B -1)s_B s_A \rangle + (J_A ^z)^2 \otimes  J_B ^z \rho_{AB}(0) (J_A ^z)^2 \otimes J_B ^z \langle (c_A -1)^2s_B ^2 \rangle \nonumber \\
& + (J_A ^z)^2 \otimes  (J_B ^z)^2 \rho_{AB}(0) \langle (c_A -1)(c_B -1) \rangle + (J_A ^z)^2 \otimes  J_B ^z \rho_{AB}(0) (J_B ^z)^2 \langle (c_A -1)(c_B -1)^2 \rangle \nonumber \\
& - (J_A ^z)^2 \otimes  J_B ^z \rho_{AB}(0) J_A ^z \otimes J_B ^z \langle (c_A -1)(c_B -1)s_A s_B \rangle + (J_A ^z)^2 \otimes  J_B ^z \rho_{AB}(0) (J_A ^z)^2 \langle (c_A -1)^2 (c_B -1)\rangle \nonumber \\
& + (J_A ^z)^2 \otimes  J_B ^z \rho_{AB}(0) (J_A ^z)^2 \otimes (J_B ^z)^2 \langle (c_A -1)^2 (c_B -1)^2 \rangle \;.
    \label{trace rhoABphi over phi delta}
\end{align}
where $\langle s_A s_B (c_A - 1)(c_B - 1) \rangle$, etc., denotes the expectation value of  
\begin{equation}
\langle s_A s_B (c_A - 1)(c_B - 1)\rangle = \langle 0| s_A s_B (c_A - 1)(c_B - 1) |0\rangle \label{cvphi} \;, 
\end{equation}
{ where 
\begin{equation} 
s_A = \frac{1}{2i}(e^{i\varphi(f_A)} - e^{-i\varphi(f_A)}) \;. \label{ws}
\end{equation} 
}
In the next section we shall see how can these correlation functions  be addressed in closed form, using the Tomita-Takesaki theory.

\section{Tomita-Takesaki modular theory and the von Neumann algebra of the Weyl Operators}\label{TTW}
\noindent To calculate the correlation functions of the Weyl operators, Eq.\eqref{cvphi}, it is worth providing a compact review of the properties of the von Neumann algebra related to such operators. For a more detailed review, one can check Refs.\cite{Weyl23,Guido:2008jk}.
\\\\ Let us begin by recalling the commutator between the scalar fields, for arbitrary spacetime separation
\begin{eqnarray}
[\varphi(x), \varphi(y)]  =  i \Delta_{PJ}(x-y),
\label{CCRscalar}
\end{eqnarray}
where the Lorentz-invariant causal Pauli-Jordan distribution $\Delta_{PJ}(x-y)$ is defined by
\begin{eqnarray}
    i \Delta_{PJ}(x-y) \! =\!\! \int \!\! \frac{d^4k}{(2\pi)^3} \varepsilon(k^0) \delta(k^2-m^2) e^{-ik(x-y)},
    \label{PJ}
\end{eqnarray}
with $\varepsilon(x) \equiv \theta(x) - \theta(-x)$. The Pauli-Jordan distribution  $\Delta_{PJ}(x-y)$ vanishes outside of the light cone, guaranteeing that measurements at points separated by space-like intervals do not interfere, that is
\begin{equation} 
\Delta_{PJ}(x-y) = 0 \;, \quad {\rm for} \quad (x-y)^2<0 \;. \label{spl}
\end{equation}
{ Now, let $\mathcal{O}$ be a subregion  of the ${\cal W}^R$} and let $\mathcal{M}(\mathcal{O})$ be the space of smooth test functions  with support contained in $\mathcal{O}$, namely
\begin{align} 
	\mathcal{M}(\mathcal{O}) = \{ f \, \vert supp(f) \subseteq \mathcal{O} \}. \label{MO}
\end{align}
Following \cite{SW1,Summers87b} one  introduces the symplectic complement  of $\mathcal{M}(\mathcal{O})$ as 
\begin{align} 
	\mathcal{M'}(\mathcal{O}) = \{ g \, \vert  \Delta_{PJ}(g,f) = 0, \; \forall f \in \mathcal{M}(\mathcal{O}) \}. \label{MpO}
\end{align}
This symplectic complement $\mathcal{M'}(\mathcal{O})$ comprises all test functions for which the smeared Pauli-Jordan expression $\Delta_{PJ}(f,g)$  vanishes for any $f$ belonging to $\mathcal{M}(\mathcal{O})$, 
\begin{equation} 
\left[ \varphi(f), \varphi(g) \right] =  i  \Delta_{PJ}(f,g) \;, \label{smpj}
\end{equation} 
allowing us to rephrase causality, Eq.\eqref{spl}, as \cite{SW1,Summers87b}
\begin{align}
    \left[ \varphi(f), \varphi(g) \right] = 0,
\end{align}
whenever $f \in \mathcal {M}(\mathcal{O})$ and $g \in \mathcal {M'}(\mathcal{O})$. \\\\As already mentioned in Sec.\eqref{intro}, the so-called Weyl operators \cite{SW1,Summers87b,Weyl23} play an important role in the study of the Bell-CHSH inequality.  This class of unitary operators is obtained by exponentiating the smeared field
\begin{equation} 
W_{h} = e^{i {\varphi}(h) }. 
\label{Weyl}
\end{equation}
By applying the Baker–Campbell–Hausdorff formula together  with the commutation relation \eqref{PJ}, one finds that the Weyl operators lead to the following algebraic structure:
\begin{align} \label{algebra} 
	W_{f} W_{g}   &= e^{ - \frac{i}{2} \Delta_{\textrm{PJ}}(f, g)}\; W_{(f+g)}, \nonumber \\
	W_{f}^{\dagger} W_{f} &= 	W_{f} W_{f}^{\dagger} = 1, \nonumber \\ 
	W^{\dagger}_{f} &=  W_{(-f)}.	 
\end{align} 
Moreover, if $f$ and $g$ are space-like, the Weyl operators $W_f$ and $W_g$ commute. By expanding the field $\varphi$ in terms of creation and annihilation operators, one can evaluate the expectation value of the Weyl operator, finding
\begin{equation} 
\langle 0 \vert  W_{h}  \vert 0 \rangle = \; e^{-\frac{1}{2} {\lVert h\rVert}^2}, 
\label{valueW}
\end{equation} 
where $\vert\vert h \vert\vert^2 = \langle h \vert h \rangle$ and 
\begin{eqnarray} 
\langle f \vert g \rangle & = & \int \frac{d^3k}{(2\pi)^3} \frac{1}{2\omega_k} f(\omega_k,\vec{k})^{*} g(\omega_k,\vec{k})\;,  \label{inner}
\end{eqnarray} 
is the Lorentz invariant inner product between the test functions $(f,g)$ \cite{SW1,Summers87b,Weyl23}, with the usual relation $\omega^2_k = {\vec k}^2+m^2$ and
\begin{align}
    f(\omega_k, \vec{k}) = \int d^4x \; e^{ikx} f(x) \;, \qquad k_0=\omega_k \;. \nonumber
\end{align}
A von Neumann algebra $\mathcal{A}(\mathcal{M})$ arises by taking all possible products and linear combinations of the Weyl operators defined on $\mathcal{M}(\mathcal{O})$. In particular, the Reeh-Schlieder theorem \cite{Haag92,Witten18,SW1,Summers87b}, states that the vacuum state $\vert 0 \rangle$ is both cyclic and separating for the von Neumann algebra $\mathcal{A}$. Consequently, we can apply the Tomita-Takesaki modular theory \cite{Bratteli97,Witten18,SW1,Summers87b,Weyl23} and introduce the anti-linear unbounded operator $S$, whose action on the von Neumann algebra $\mathcal{A}(\mathcal{M})$ is defined as
\begin{align} 
	S \; a \vert 0 \rangle = a^{\dagger} \vert 0 \rangle, \qquad \forall a \in \mathcal{A}(\mathcal{M}),
    \label{TT1}
\end{align}  
from which it follows that $S^2 = 1$ and $S \vert 0 \rangle = \vert 0 \rangle$. The operator $S$ has a unique polar decomposition \cite{Bratteli97}:
\begin{align}
    S = J  \Delta^{1/2},
    \label{PD}    
\end{align} 
where $J$ is anti-unitary  and $\Delta$ is positive and self-adjoint. These  operators are characterized by the following set of properties \cite{Bratteli97,Witten18,SW1,Summers87b,Weyl23}:
\begin{eqnarray}
	\Delta & = & S^{\dagger} S \;, \quad J \Delta^{1/2} J = \Delta^{-1/2} \;, \nonumber \\
	J^2 & = & 1 \;, \quad S^{\dagger} = J \Delta^{-1/2} \;, \nonumber \\
	J^{\dagger} & = & J \;,  \quad \Delta^{-1} = S S^{\dagger} \;.
    \label{TTP}
\end{eqnarray}
From the Tomita-Takesaki theorem \cite{Bratteli97,Witten18,SW1,Summers87b,Weyl23}, it follows that $J \mathcal{A}(\mathcal{M}) J = \mathcal{A}'(\mathcal{M})$, meaning that, upon conjugation by the operator $J$, the algebra $\mathcal{A}(\mathcal{M})$ is mapped onto its commutant $\mathcal{A'}(\mathcal{M})$, namely: 
\begin{equation} 
\mathcal{A'}(\mathcal{M}) = \{ \; a' \, \vert \; [a,a']=0, \forall a \in \mathcal{A}(\mathcal{M}) \;\}.
    \label{commA}
\end{equation}
Furthermore, the theorem states that there is a one-parameter family of operators $\Delta^{it}$, $t \in \mathbb{R}$, that leave the algebra $\mathcal{A}(\mathcal{M})$ invariant, such that the the following equation holds
\begin{align*}
    \Delta^{it} \mathcal{A}(\mathcal{M}) \Delta^{-it} = \mathcal{A}(\mathcal{M}) \;.
\end{align*}
The Tomita-Takesaki modular theory is particularly well-suited for analyzing the Bell-CHSH inequality within the framework of relativistic Quantum Field Theory \cite{SW1,Summers87b}. As demonstrated in \cite{Weyl23}, it provides a purely algebraic method for constructing Bob's operators from Alice's ones by using the modular conjugation $J$. Given Alice's operator $A_f$, one can assign to Bob the operator $B_f = J A_f J$, ensuring their mutual commutativity due to the Tomita-Takesaki theorem, as $B_f = J A_f J$ belongs to the commutant $\mathcal{A'}(\mathcal{M})$ \cite{Weyl23}. \\\\An important outcome of the Tomita-Takesaki modular theory, established by  \cite{Rieffel77,Eckmann73}, allows the extension of the action of the modular operators $(J, \Delta)$ to the space of the test functions. In fact, when equipped with the Lorentz-invariant inner product $\langle f \vert g\rangle$, Eq.\eqref{inner}, the set of test functions forms a complex Hilbert space $\mathcal{F}$ that possesses a variety of properties. To be more precise, it is found that the subspaces $\mathcal{M}$ and $i\mathcal{M}$ are standard subspaces for $\mathcal{F}$ \cite{Rieffel77}. This implies that:
\begin{enumerate}[label=\roman*.]
\item $\mathcal{M} \cap i \mathcal{M} = \{ 0 \}$;
\item $\mathcal{M} + i \mathcal{M}$ is dense in $\mathcal{F}$. 
\end{enumerate}
As shown in  \cite{Rieffel77}, for such subspaces,  it's viable to establish a modular theory similar to that of the Tomita-Takesaki theory. This involves introducing an operator $s$ acting on $\mathcal{M} + i\mathcal{M}$ such that
\begin{align}
    s (f+ih) = f-ih \;, 
    \label{saction}
\end{align}
for $f,h \in \mathcal{M}$. With this definition, it's worth noting that $s^2 = 1$. Employing the  polar decomposition, one obtains:  
\begin{align}
    s = j \delta^{1/2},
\end{align}
where $j$ is an anti-unitary operator, while $\delta$ is positive and self-adjoint.  Similarly to the operators $(J,\, \Delta)$, the  operators $(j,\, \delta)$ fulfill  the following properties \cite{Rieffel77}:
\begin{align}
    j \delta^{1/2} j &= \delta^{-1/2}\;, \,\,\,\,\,\,  \delta^\dagger = \delta\;,\nonumber \\
    s^\dagger &= j \delta^{-1/2}\;, \,\,\, j^\dagger = j\,; \nonumber \\
    \delta &= s^\dagger s\;, \,\,\,\,\,\,\,\,\,\,\, j^2=1 \;.
\end{align}
Further, one can show \cite{Summers87b,Rieffel77} that a test function $f$ belongs to $\mathcal{M}$ if and only if 
\begin{equation} 
s f = f \;. 
    \label{sff}
\end{equation}
Indeed, let us suppose that $f \in \mathcal{M}$.  From Eq.\eqref{saction}, one can express 
\begin{equation}
sf = h_1 + i h_2 \;, 
    \label{pv1}
\end{equation}
for some $(h_1,h_2)$. Since $s^2=1$ it follows that 
\begin{equation} 
f = s(h_1 + i h_2) = h_1 -i h_2  \;, \label{pv2}
\end{equation} 
so that $h_1=f$ and $h_2=0$. Similarly, one has that $f' \in \mathcal{M}'$ if and only if $s^{\dagger} f'= f'$. \\\\Thus, the lifting of the action of the operators $(J, \Delta)$ to the space of test functions is accomplished by \cite{Eckmann73} 
\begin{align} 
 J e^{i {\varphi}(f) } J  = e^{-i {\varphi}(jf) }, \quad \Delta e^{i {\varphi}(f) } \Delta^{-1} = e^{i {\varphi}(\delta f) }. \label{jop}
\end{align} 
Also, it is important to note that if $f \in \mathcal{M} \implies jf \in \mathcal{M}'$. This property follows from 
\begin{equation} 
s^{\dagger} (jf) = j \delta^{-1/2} jf = \delta f = j (j\delta f) = j (sf) = jf \;. \label{jjf} 
\end{equation} 
{ It is worth reminding here that  $\delta$  is an unbounded operator with continuous spectrum. For instance, as one learns from the work of \cite{Bisognano75}, for the wedge ${\cal W}^R$, the spectrum of $\delta$ coincides with the positive real line, {\it i.e.}, $\log(\delta) = \mathbb{R}$. In the case of a continuous spectrum we lack the notion of eigenstates. Rather, it is appropriate to make use of the spectral decomposition \cite{Bratteli97} of the operator $\delta$ and refer to spectral subspaces ${\sigma}_{\lambda}$, parametrized by a real parameter $\lambda \in \mathbb{R}_{+}$.}\\\\We  have now all the necessary ingredients to evaluate the correlation functions of the Weyl operators. By examining expression \eqref{trace rhoABphi over phi delta}, one recognizes  that the fundamental quantity to be computed is of the form
\begin{equation}
\langle e^{i \varphi(f_A)} e^{ \pm i\varphi(f_B)}\rangle = \langle e^{i( \phi(f_A)\pm \phi(f_B))}\rangle = e^{-\frac{1}{2} ||f_A \pm f_B||^2} \;, \label{Wex}
\end{equation}
so that  we need to evaluate  the following norms $(||f_A||^2, ||f_B||^2) $ and the inner product $\langle f_A | f_B \rangle$. We begin by focusing on Alice’s test function $f_A$. We require that $f_A \in {\cal M(O)}$ where ${\cal O}$ is located in the right Rindler wedge. Following \cite{SW1,Summers87b,Weyl23}, the test function $f_A$ can be further specified by considering the spectrum of the operator $\delta$. By selecting the subspace ${\sigma}_{\lambda}=[\lambda^2-\varepsilon, \lambda^2+ \varepsilon]$  and introducing the normalized vector $\phi$ belonging to this subspace,  one writes 
\begin{equation}
f _A = \eta  (1+s) \phi \;,
\label{nmf}
\end{equation}
where $\eta$ is an arbitrary parameter. As required by the setup outlined above, Eq.\eqref{nmf} ensures that 
\begin{equation}
s f_A = f_A  \;. \label{fafa}
\end{equation}
We observe that $j\phi$ is orthogonal to $\phi$, {\it i.e.}, $\langle \phi |  j\phi \rangle = 0$. In fact, from 
\begin{align} 
\delta^{-1} (j \phi) =  j (j \delta^{-1} j) \phi = j (\delta \phi), 
\label{orth}
\end{align}
it follows that the modular conjugation $j$ exchanges the spectral subspace $[\lambda^2-\varepsilon, \lambda^2+\varepsilon ]$ with $[1/\lambda^2-\varepsilon,1/ \lambda^2+\varepsilon ]$. Regarding Bob's test function $f_B$, we use the modular conjugation operator $j$ and define 
\begin{equation} 
f_B = j f_A \;, \label{fb}
\end{equation}
ensuring that 
\begin{equation} 
s^{\dagger} f_B = f_B
\end{equation}
This implies that, as required by the relativistic causality, $f_B$ belongs to the symplectic complement $\mathcal{M'}(\mathcal{O})$, located in the left Rindler wedge, namely: $f_B \in  \mathcal{M'}(\mathcal{O})$. Finally, considering that $\phi$ belongs to the spectral subspace $[\lambda^2-\varepsilon, \lambda^2+\varepsilon ] $, it follows that \cite{Weyl23}, 
\begin{align}
\vert\vert f_A \vert\vert^2  &= \vert\vert jf _A \vert\vert^2 = \eta^2 (1+\lambda^2) \;, \nonumber \\
\langle f_A \vert jf_A \rangle &= 2 \eta^2 \lambda  \;, \label{sfl}
\end{align}
which provide us the needed inner products. 

\section{The Bell-CHSH inequality}\label{BCHSH}

\noindent We face now the Bell-CHSH inequality, Eq.\eqref{tracerhoC}. We begin by defining the Bell operators \cite{SW1,Summers87b,Sorella:2023iwz}: 
\begin{align}     
    A \vert -1 \rangle & = e^{i \alpha} \vert 1 \rangle \;, \quad &
    A \vert 0 \rangle & = \vert 0 \rangle \;, \quad &
    A \vert 1 \rangle & = e^{-i \alpha} \vert -1 \rangle \nonumber \\
    A' \vert -1 \rangle & = e^{i \alpha'} \vert 1 \rangle \;, \quad &
    A' \vert 0 \rangle & = \vert 0 \rangle \;, \quad &
    A' \vert 1 \rangle & = e^{-i \alpha'} \vert -1 \rangle \nonumber \\
    B \vert -1 \rangle & = e^{-i \beta} \vert 1 \rangle \;, \quad &
    B \vert 0 \rangle & = \vert 0 \rangle \;, \quad &
    B \vert 1 \rangle & = e^{i \beta} \vert -1 \rangle \nonumber \\
    B' \vert -1 \rangle & = e^{-i \beta'} \vert 1 \rangle \;, \quad & 
    B' \vert 0 \rangle & = \vert 0 \rangle \;, \quad &
    B' \vert 1 \rangle & = e^{i \beta'} \vert -1 \rangle \;, \label{bopp}
\end{align}
which fulfill the whole set of conditions \eqref{Bop}. The free parameters $(\alpha, \alpha',\beta, \beta')$, which will be chosen at the best convenience, correspond to the four Bell's angles.  \\\\Reminding that the initial state for $AB$ is 
\begin{align*}
    \vert \psi \rangle_{AB} = \frac{1}{\sqrt{3}} \left ( \vert -1 \rangle_{A} \otimes\vert -1 \rangle_{B} - \vert 0 \rangle_{A} \otimes\vert 0 \rangle_{B} + \vert 1 \rangle_{A} \otimes\vert 1 \rangle_{B} \right ),
\end{align*}
and using Eq.\eqref{tracerhoC}, {  one gets the Bell-CHSH correlator
\begin{align}
    \langle \mathcal{C} \rangle &= \frac{1}{3}[1+2 \cos(\alpha + \beta)] + \frac{2}{3}\cos(\alpha+\beta)[2 \langle(c_A-1)\rangle + 2\langle(c_B-1)\rangle + 4\langle s_A s_B \rangle \nonumber \\
    &\quad - \langle s_A ^2 \rangle - \langle s_B ^2 \rangle + 4 \langle s_A s_B (c_B - 1) \rangle + 4 \langle s_A s_B (c_A - 1) \rangle - 2 \langle s_B ^2 (c_A - 1) \rangle \nonumber \\
    &\quad - 2 \langle s_A ^2 (c_B - 1) \rangle + 4\langle (c_A -1)(c_B -1) \rangle + \langle (c_A -1)^2 \rangle + \langle (c_B -1)^2 \rangle \nonumber \\
    &\quad + 2 \langle (c_A -1)(c_B -1)^2 \rangle + 2 \langle (c_A -1)^2(c_B -1) \rangle + \langle s_A ^2 s_B ^2 \rangle \nonumber \\
    &\quad + 4 \langle s_A s_B (c_A -1)(c_B -1) - \langle s_A ^2 (c_B -1)^2 \rangle - \langle s_B ^2 (c_A -1)^2 \rangle
    \nonumber \\ &\quad + \langle (c_A -1)^2 (c_B -1)^2 \rangle]  \nonumber \\
    & + \frac{1}{3}[1+2 \cos(\alpha' + \beta)] + \frac{2}{3}\cos(\alpha+\beta)[2 \langle(c_A-1)\rangle + 2\langle(c_B-1)\rangle + 4\langle s_A s_B \rangle \nonumber \\
    &\quad - \langle s_A ^2 \rangle - \langle s_B ^2 \rangle + 4 \langle s_A s_B (c_B - 1) \rangle + 4 \langle s_A s_B (c_A - 1) \rangle - 2 \langle s_B ^2 (c_A - 1) \rangle \nonumber \\
    &\quad - 2 \langle s_A ^2 (c_B - 1) \rangle + 4\langle (c_A -1)(c_B -1) \rangle + \langle (c_A -1)^2 \rangle + \langle (c_B -1)^2 \rangle \nonumber \\
    &\quad + 2 \langle (c_A -1)(c_B -1)^2 \rangle + 2 \langle (c_A -1)^2(c_B -1) \rangle + \langle s_A ^2 s_B ^2 \rangle \nonumber \\
    &\quad + 4 \langle s_A s_B (c_A -1)(c_B -1) - \langle s_A ^2 (c_B -1)^2 \rangle - \langle s_B ^2 (c_A -1)^2 \rangle
    \nonumber \\ &\quad + \langle (c_A -1)^2 (c_B -1)^2 \rangle] \nonumber \\
     & + \frac{1}{3}[1+2 \cos(\alpha + \beta')] + \frac{2}{3}\cos(\alpha+\beta')[2 \langle(c_A-1)\rangle + 2\langle(c_B-1)\rangle + 4\langle s_A s_B \rangle \nonumber \\
    &\quad - \langle s_A ^2 \rangle - \langle s_B ^2 \rangle + 4 \langle s_A s_B (c_B - 1) \rangle + 4 \langle s_A s_B (c_A - 1) \rangle - 2 \langle s_B ^2 (c_A - 1) \rangle \nonumber \\
    &\quad - 2 \langle s_A ^2 (c_B - 1) \rangle + 4\langle (c_A -1)(c_B -1) \rangle + \langle (c_A -1)^2 \rangle + \langle (c_B -1)^2 \rangle \nonumber \\
    &\quad + 2 \langle (c_A -1)(c_B -1)^2 \rangle + 2 \langle (c_A -1)^2(c_B -1) \rangle + \langle s_A ^2 s_B ^2 \rangle \nonumber \\
    &\quad + 4 \langle s_A s_B (c_A -1)(c_B -1) - \langle s_A ^2 (c_B -1)^2 \rangle - \langle s_B ^2 (c_A -1)^2 \rangle
    \nonumber \\ &\quad + \langle (c_A -1)^2 (c_B -1)^2 \rangle] \nonumber \\
 & - \frac{1}{3}[1+2 \cos(\alpha' + \beta')] - \frac{2}{3}\cos(\alpha'+\beta')[2 \langle(c_A-1)\rangle + 2\langle(c_B-1)\rangle + 4\langle s_A s_B \rangle \nonumber \\
    &\quad - \langle s_A ^2 \rangle - \langle s_B ^2 \rangle + 4 \langle s_A s_B (c_B - 1) \rangle + 4 \langle s_A s_B (c_A - 1) \rangle - 2 \langle s_B ^2 (c_A - 1) \rangle \nonumber \\
    &\quad - 2 \langle s_A ^2 (c_B - 1) \rangle + 4\langle (c_A -1)(c_B -1) \rangle + \langle (c_A -1)^2 \rangle + \langle (c_B -1)^2 \rangle \nonumber \\
    &\quad + 2 \langle (c_A -1)(c_B -1)^2 \rangle + 2 \langle (c_A -1)^2(c_B -1) \rangle + \langle s_A ^2 s_B ^2 \rangle \nonumber \\
    &\quad + 4 \langle s_A s_B (c_A -1)(c_B -1) - \langle s_A ^2 (c_B -1)^2 \rangle - \langle s_B ^2 (c_A -1)^2 \rangle
    \nonumber \\ &\quad + \langle (c_A -1)^2 (c_B -1)^2 \rangle] \;. \label{ccc}
\end{align}
}
The expression above is written in terms of the inner products between test functions, which can be evaluated by employing the expressions \eqref{sfl}. The final expression reads
\begin{align}
    \langle \mathcal{C} \rangle &= \frac{2}{3} \left \{ 1+ [\cos(\alpha+\beta) + \cos(\alpha'+\beta)+\cos(\alpha+\beta')-\cos(\alpha'+\beta')] \right \} \nonumber \\ &\quad -\frac{4}{3} f(\eta, \lambda) [\cos(\alpha +\beta) + \cos(\alpha'+\beta)+\cos(\alpha+\beta')-\cos(\alpha'+\beta')]\;,
    \label{ffb}
\end{align}
where the function $f(\eta,\lambda)$ is
\begin{align}
    f(\eta,\lambda) = e^{-2\eta^2(1+\lambda^2)} - e^{-4\eta^2(1-\lambda)^2}\;,
    \label{fetalambda}
\end{align}
with $\eta \neq 0$. 
From \eqref{ffb}, one learns several things:
\begin{itemize}
\item When the quantum field $\varphi$ is removed, {\it i.e.} $\eta^2 \rightarrow 0$, and therefore, $f(\eta,\lambda) = 0$, we recover the Bell-CHSH inequality of Quantum Mechanics for  qutrits, whose maximum value  is \cite{Gisin,Peruzzo:2023nrr}
\begin{align}
    \langle \mathcal{C} \rangle_{f=0} = \frac{2}{3} (1+2\sqrt{2}) \approx 2.55 > 2.
\end{align}
One  notices that this value is lower than  Tsirelson's bound, as we are dealing with a spin $1$ system.

\item The contributions arising  from  the scalar field $\varphi$ are encoded in the exponential terms  $e^{-4 \eta^2(1-\lambda)^2}$ and $e^{-2\eta^2(1+\lambda^2)}$. It is worth reminding here that the parameter $\eta^2$ is related to the norm of the test function $f_A$, Eqs.\eqref{sfl}, that is, this parameter reflects the freedom one has in defining the test function $f_A$ through the operator $s$.  As pointed out in \cite{Weyl23,Guimaraes:2024alk}, $\eta$ is a free parameter appearing in the Quantum Field Theory formulation of the Bell-CHSH inequality in terms of Weyl operators, playing a similar role of the free Bell's angles and it can bee chosen in the most suitable way. {This feature can be understood as follows. Looking at the Bell's operators $(A,A',B,B')$, Eqs.\eqref{bopp}, one realizes that they are dichotomic for arbitrary values of the parameters $(\alpha, \alpha', \beta, \beta')$. As such, they  are completely free and, in fact, are chosen at the best convenience in the final expression of the Bell-CHSH inequality. The same pattern is encountered in the case of the parameter $\eta$. One has to notice that the Weyl operator 
\begin{equation} 
W_{f_A}= e^{i \varphi(f_A)} = e^{i \varphi(\eta (1+s)\phi) }\;, \label{wfa}
\end{equation} 
is unitary for any value of the parameter $\eta$. }

\item   { We have now to face the choice of the spectral subspace ${\sigma}_{\lambda}$ of the modular operator $\delta$. This is a not easy task due to the fact that $\delta$ has a continuous spectrum given by the positive real line $\mathbb{R}_{+}$. For a better illustration of this point, we remind here the expression found in \cite{Summers87b} for the violation of the Bell-CHSH in the vacuum state of a quantum scalar field\footnote{See Sec.III, just before corollary 3.2.}, namely 
\begin{equation} 
\frac{4 \sqrt{2} \lambda}{1+ \lambda^2} \;, \label{sw}
\end{equation}
from which the choice of the spectral subspace $[\sqrt{2}-1, \sqrt{2}+1 ] $  follows. The maximum violation is attained for $\lambda =1$. The important point here is that this choice can be made, as $\lambda=1$ belongs to the continuous spectrum of $\delta$. \\\\In our case we proceed as follows. From $e^{-2\eta^2(1+\lambda^2)} > e^{-4\eta^2(1-\lambda^2)}$, we get the roots $\lambda\pm = 2 \pm \sqrt{3}$. We can distinguish two possibilities. The first one is when $0 < \lambda < 2-{\sqrt{3}}=0.267$. In this case, the quantum field produces a damping, resulting in a decreasing of the violation of the Bell-CHSH inequality, as compared to the pure Quantum Mechanical  case. The second possibility takes place  when $ 2-\sqrt{3}< \lambda < 0.3$, resulting in an improvement of the size of the violation. In other words, we pick up the spectral subspace $[0,0.3]$. This spectral subspace has been identified through a numerical analysis. Again, it is an allowed choice. }
\item { Let us also check that the Tsirelson bound is fulfilled for arbitrary values of the parameter $\eta$.  Observing that the  maximum value of the angular part of Eq.\eqref{ffb} is $2 \sqrt{2}$, attained for the following values of the Bell's angles:
\begin{eqnarray}
    \alpha & = & 0 \;, \quad \alpha' = \frac{\pi}{2} \;,	\quad \beta = -\frac{\pi}{4} \;, \quad \beta' = \frac{\pi}{4}\;,
    \label{bellanglesmax}
\end{eqnarray}
it follows that
\begin{align}
    \langle \mathcal{C} \rangle = \frac{2}{3}\{1+2\sqrt{2}[1-2f(\eta,\lambda)]\}.
\end{align}
Accordingly, Tsirelson bound,  $\langle \mathcal{C} \rangle \leq 2\sqrt{2}$. is fulfilled whenever 
 \begin{align}
    f(\eta,\lambda) \ge \frac{1-\sqrt{2}}{4 \sqrt{2}} \;, 
    \label{fcondition}
\end{align}
with $\lambda \in [0.0.3]$. A numerical investigation shows that eq.\eqref{fcondition} is in fact fulfilled for arbitrary values of the normalization factor $0<\eta <\infty$\footnote{Since the function $f(\eta, \lambda)$ depends only from $\eta^2$, we can always take $\eta$ to be non-negative.}. see, e.g.  Fig.\eqref{fff} for the behavior of $f(\eta,\lambda)$ for $\eta \in [0,10]$. 

\begin{figure}[h]
\includegraphics[width=7cm]{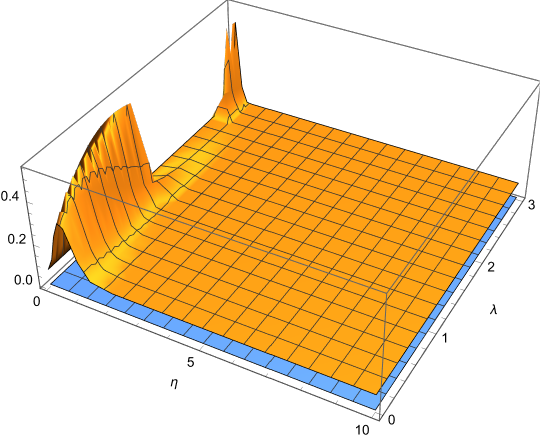}
\caption{The orange surface shows the behavior of  $f(\eta,\lambda) - \frac{1-\sqrt{2}}{4 \sqrt{2}}$ for $\eta \in [0, 10]$. The blue surface is the plane
correponding to $f(\eta,\lambda) - \frac{1-\sqrt{2}}{4 \sqrt{2}}=0 $. One sees that the orange surface is always above the blue plane, showing that Tsirelson's bound is fulfilled. Similar plots are obtained when increasing $\eta$. }
	\label{fff}
	\end{figure}
}

\item The whole effects produced by the quantum field can be captured in Fig. \eqref{Fig plot3d}.  The orange surface represents the maximum value of $\langle \mathcal{C} \rangle$ without the presence of $\varphi$, \textit{i.e.} $\langle \mathcal{C} \rangle = 2.55$. One  notices the existence of a small region in blue, above the orange surface. This region corresponds to values of $(\eta, \lambda)$ for which the size of the violation is improved, almost till approximately $2.7$. This phenomenon occurs when $0.3 > \lambda > 0.26$.
\begin{figure}[h]
\includegraphics[width=8cm]{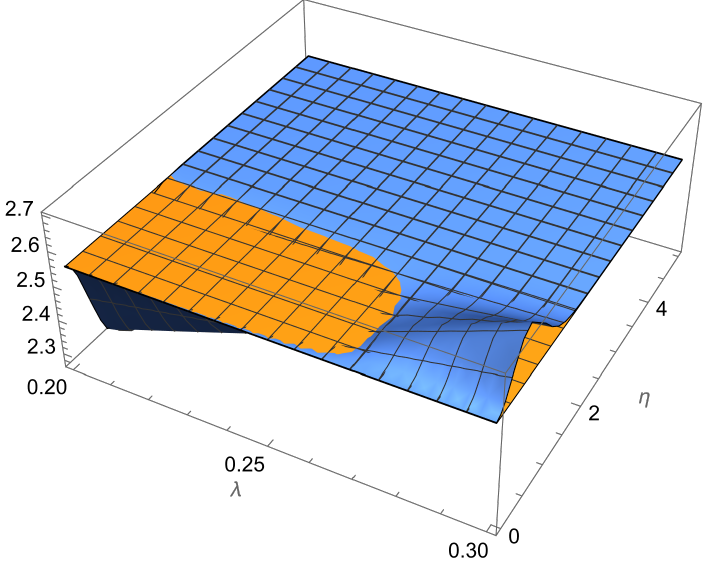}
\caption{Behavior of the Bell-CHSH correlator $\langle \mathcal{C} \rangle$ as a function of the parameters $\eta$ and $\lambda$. The orange surface represents the maximum value of $\langle \mathcal{C} \rangle$ without the presence of $\varphi$, \textit{i.e.} $\langle \mathcal{C} \rangle = 2.55$.The blue surface above the orange one corresponds to the region in which the effects of the quantum field result in an increasing of the size of the violation of the Bell-CHSH inequality.}
	\label{Fig plot3d}
	\end{figure}
	\end{itemize}

{
\subsection{Comparison  with the spin $1/2$ case} 
\noindent Let us end this section by reminding the results obtained in the case of spin $1/2$, in the dephasing channel, as reported in \cite{spinhalf24}. \\\\For the initial state we have 
\begin{equation} 
\left( \frac{|+\rangle_A \otimes |+\rangle_B + |-\rangle_A \otimes |-\rangle_B}{\sqrt{2}} \right)\otimes |0\rangle  \;. \label{12}
\end{equation}
The Bell-CHSH inequality is found 
\begin{equation} 
\langle {\cal C} \rangle_{1/2} = 2 \sqrt{2} \;e^{-\eta^2 (1+\lambda)^2} \;, \label{b12}
\end{equation} 
which, unlike the case of the spin $1$, exhibits only a decreasing of the size of the violation. 
}

\section{Conclusions} \label{conclusion}
\noindent In this work, we have analyzed the interaction between spin $1$ Unruh-De Witt detectors, \textit{i.e.} a pair of qutrits, and a relativistic quantum scalar field $\varphi$. The effects of the quantum field on the Bell-CHSH inequality have been scrutinized in detail by making use of the dephasing channel for the evolution operator. By employing the Tomita-Takesaki modular theory and the properties of the Weyl operators, these effects have been evaluated in closed form, as expressed by Eq.\eqref{ffb}. \\\\The main finding of the present study is that the presence of a scalar quantum field may induce both a damping as well as an improvement effect, resulting, respectively,  in a decreasing and an increasing of the size of the violation of the Bell-CHSH inequality as compared to the case in which the field is absent. \\\\As such, the case of spin $1$ looks much different from that of spin $1/2$, for which only a decreasing of the violation has been detected \cite{spinhalf24}. As already underlined, the existence of an improvement of the size of the violation of the Bell-CHSH inequality can be ascribed to the fact that, for spin $1$, the Tsirelson bound $2\sqrt{2}$ is never saturated. Instead, the maximum value obtained in Quantum Mechanics is $[\frac{2}{3}(1+ 2\sqrt{2})\approx 2.55$. As such, in the presence of a quantum field $\varphi$, there exists a  permissible interval, $[\frac{2}{3}(1+ 2\sqrt{2}), 2\sqrt{2}]$, where an increasing of the size of the violation occurs.

\section*{Acknowledgments}
\noindent	The authors would like to thank the Brazilian agencies CNPq and FAPERJ, for financial support.  S.P.~Sorella, I.~Roditi, and M.S.~Guimaraes are CNPq researchers under contracts 301030/2019-7, 311876/2021-8, and 309793/2023-8, respectively. F.M.~Guedes acknowledges FAPERJ for financial support under the contract SEI-260003/007871/2024.	
\
	
\end{document}